\documentclass[twoside,11pt]{article}

\usepackage{graphicx}

\usepackage{amssymb}
\usepackage{amsmath}
\usepackage{amsfonts}
\usepackage{wasysym}
\usepackage{revsymb}
\usepackage{amsbsy}
\usepackage{bm}

\usepackage{titlesec}

\usepackage{cite}

\titleformat*{\section}{\bfseries}
\titleformat*{\subsection}{\bfseries}

\begin{document}           

\title{\Large 
Minisuperspace model of quantum geometrodynamics in the Madelung-Bohm formalism}
\author{V.E. Kuzmichev, V.V. Kuzmichev\\[0.5cm]
\itshape Bogolyubov Institute for Theoretical Physics,\\
\itshape National Academy of Sciences of Ukraine, Kyiv, 03143 Ukraine}

\date{}

\maketitle

\begin{abstract}
An analogy between non-relativistic quantum mechanics in the Madelung formulation and quantum geometrodynamics in the case of the maximally symmetric 
space is drawn. The equations equivalent to the continuity equation and the hydrodynamic Euler equation describing the evolution of the velocity introduced for 
the case of hypothetical fluid flow characterizing the cosmological system are obtained. It is shown that the perfect nature of the fluid is broken by the quantum 
Bohm potential. The quantum potential is calculated in semi-classical approximation for different forces acting in the system both in standard quantum 
mechanics and in minisuperspace model of quantum geometrodynamics. The explicit dependences of the cosmic scale factor on the conformal time, which 
take into account the quantum additive, are found for empty space with spatial curvature and for a spatially flat universe with dust and radiation.
\end{abstract}


\section{Introduction}\label{sec:1}
Madelung's paper \cite{Ma27} on quantum theory in hydrodynamical form can be viewed as one of the most seminal in the scope of quantum theory. It became 
the starting point for many studies on the relationship between the newly emerged quantum theory based on the Schr\"{o}dinger equation \cite{Sch26} and the 
well-established classical theories of mechanics and electrodynamics, as well as for the attempts to interpret quantum theory itself. There was an intense 
discussion about the issue and the necessity of finding a consistent formulation of quantum theory in the 1950s \cite{B52,B52b,T52,H52,B52c,B53,B54,Fr57}. 
Madelung's formalism played an important role in introducing the idea of an ensemble of classical trajectories in the further development of quantum 
mechanics. Specifically, Dirac's idea  \cite{Dir58}, realized in Feynman's path integral formulation of quantum mechanics, can be discerned in Madelung's 
hydrodynamic approach.

In the last two decades, there has been renewed interest in the Madelung–Bohm representation of wavefunctions \cite{Be24}. The elements of the Madelung 
approach can be seen in the quantum Hamilton-Jacobi formalism \cite{Le83,Le83b} which provides an alternative approach to non-relativistic quantum 
mechanics and can be used, in particular, in computational method for one-dimensional scattering problems \cite{Ch06,Ch06b}. The three-dimensional 
quantum Hamilton-Jacobi equation in the case of separated variables describe physical states not detected by the Schr\"{o}dinger wave function 
\cite{Bo06,Bo09}.

Classical mechanics is widely believed to emerge from quantum mechanics as its limiting case \cite{R64}. The question of the correct transition from the 
quantum description to the classical one is actual and still open. The resolution of this question may appear to be more feasible in the context of the 
Madelung--Bohm approach \cite{Bow05}. On the other hand, this approach allows one to describe the interaction of classical and quantum
systems, when classical motion is not regarded as an approximation of quantum mechanics \cite{GB19}.

Since the Madelung equations transform the non-relativistic time-dependent Schr\"{o}dinger equation into hydrodynamic equations of non-classical virtual 
Eulerian fluid \cite{So91}, one can analyze the properties of the Madelung fluid, such as entropy \cite{He16}, and establish the relation between the Fisher 
information and the thermodynamic-like internal energy of this fluid \cite{He15}.

The exact mathematical relationship between the Madelung equations and the Schr\"{o}\-din\-ger equation is still a matter of debate \cite{Re17,Re23}. It was 
argued that consistent mathematical theory for this system of equations cannot be developed without the imposition of an additional quantization condition \cite{T52,Wa94,Dir31}. Some 
approaches claim that the Madelung equations are physically more fundamental than the Schr\"{o}dinger equation. However, if we consider the Madelung 
equations as an `unrolling' of the Schr\"{o}dinger equation into the form, in which the contribution of quantum effects is explicitly singled out (without 
considering the reverse procedure of recovering the Schr\"{o}dinger equation), then certain questions about the 
status of the Madelung equations disappear. The proportionality of quantum effects to $\hbar^{2}$ makes it possible to develop the WKB method.

The objective of this paper is not to provide a comprehensive review of all existing studies in this area of research. As mentioned above, the Madelung--Bohm 
approach remains a guideline for many current studies in quantum theory, in particular with regard to its connection with classical theory. The brief description 
of this approach given in Sect.~\ref{sec:2} is intended to present some of its key points, with a view to further application in quantum geometrodynamics. In 
order to find out the specific dependences of a quantum Bohm potential on a spatial variable, several particular cases of potentials in which a quantum particle 
moves are considered. When studying the properties of the Bohm potential, we restrict ourselves to the semi-classical approximation with respect to 
$\hbar^{2}$.  Then, the Bohm potential can be calculated directly from the potential field in which the particle moves. In Sect.~\ref{sec:3}, we consider the 
homogeneous and isotropic quantum cosmological system (universe). In this case, we can draw direct parallels with quantum theory in $(1+1)$ dimensions. 
The equations of quantum geometrodynamics for a system filled with a homogeneous scalar field and a perfect fluid are reduced to a set of equations of 
hydrodynamic form. Analogs of the quantum Bohm potential for the cases of radiation dominance, matter dominance, for a cosmological model with spatially 
flat geometry filled with radiation and dust, and for empty space with spatial curvature are calculated in the semi-classical approximation. The impact of the 
quantum potential on the expansion of the universe is studied here. Sect.~\ref{sec:4} discusses the results obtained in the previous sections.

\section{Madelung--Bohm approach in $(1+1)$ dimensions}\label{sec:2}
\subsection{Basic equations}\label{sec:2.1}
The reason for choosing for consideration $(1+1)$ dimensions is induced by the observation that quantum geometrodynamics for maximally symmetric 
spacetime reduces to a problem in $(1+1)$ dimensions. This allows us to draw parallels between quantum mechanics and quantum cosmology directly. 

In non-relativistic quantum theory, the basic equations \cite{Ma27,Sch26}
\begin{equation}\label{1}
\Psi = \psi \, e^{-\frac{i}{\hbar} W t},
\end{equation}
\begin{equation}\label{2}
- \frac{\hbar^{2}}{2 m}\, \partial _{x}^{2} \psi + \left[V(x) - W \right] \psi = 0
\end{equation}
describe the motion of a particle of mass $m$ along the $x$-axis under the action of a potential $V(x)$. Here $\psi = \psi(x)$, $\Psi = \Psi(x,t)$, 
$t$ is time, and $W$ is a total energy. Since $\Psi$ depends on $t$ linearly, Eqs.~(\ref{1}) and (\ref{2}) may be rewritten as a single equation,
\begin{equation}\label{3}
i \hbar \partial _{t} \Psi = \left[- \frac{\hbar^{2}}{2 m}\, \partial _{x}^{2} + V(x) \right]\Psi.
\end{equation}
This equation is known as the time-dependent Schr\"{o}dinger equation. We will refrain from generalizing Eq.~(\ref{3}) to the case when the potential function 
depends on time, because then it has no solution in the form (\ref{1}).

The continuity equation follows from Eq.~(\ref{3}),
\begin{equation}\label{4}
\partial _{t} \rho + \partial _{x} J = 0,
\end{equation}
where $\rho = |\Psi|^{2}$ is the probability density for finding a particle at some point in space, and 
\begin{equation}\label{5}
J = \frac{\hbar}{2 i m} \left(\Psi^{*} \partial _{x} \Psi - \Psi \partial _{x} \Psi^{*} \right)
\end{equation}
is the density of the probability current of a particle at point $x$ and at time $t$.

We will look for the solution of Eq.~(\ref{3}) in polar form
\begin{equation}\label{6}
\Psi (x, t) = A(x, t)\,e^{\frac{i}{\hbar} S(x, t)},
\end{equation}
where the amplitude $A$ and the phase $S$ are real functions of their arguments. Substituting Eq.~(\ref{6}) into Eq.~(\ref{3}) yields two equations \cite{Ma27},
\begin{equation}\label{7}
m\, \partial _{t} A^{2} + \partial _{x} \left(A^{2} \partial _{x} S \right) = 0,
\end{equation}
\begin{equation}\label{8}
\partial _{t} S + \frac{1}{2 m} (\partial _{x} S)^{2} + V = Q,
\end{equation}
where
\begin{equation}\label{9}
Q = \frac{\hbar^{2}}{2 m} \frac{\partial_{x}^{2} A}{A}
\end{equation}
is the quantum Bohm potential\footnote{This potential is usually defined with a minus sign. We leave the plus sign as it appears when deriving Eq.~(\ref{8}).} 
\cite{B52}. The potential $Q$ can be considered as an additional source 
generated by variations in amplitude $A$ in space. In the formal limit 
$\hbar^{2} \rightarrow 0$, we have $Q = 0$, and Eq.~(\ref{8}) in this limit resembles the Hamilton--Jacobi equation, if the phase $S$ is viewed as a classical action \cite{Dir58,Lan65}. By analogy with classical mechanics, we define the velocity $u$ by the 
expression
\begin{equation}\label{10}
u = \frac{\partial _{x} S}{m}.
\end{equation}
Substitution of Eq.~(\ref{6}) into Eq.~(\ref{5}) gives $J = \rho u$, and Eq.~(\ref{4}) turns into a continuity equation as in hydrodynamics,
\begin{equation}\label{11}
\partial _{t} \rho + \partial _{x} (\rho u) = 0.
\end{equation}
This equation is Eq.~(\ref{7}) in new notations. It describes the conservation law for a fluid of density $\rho$ and current $\rho u$. 

Equation (\ref{8}) can be interpreted as the quantum Hamilton-Jacobi equation, although there have been discussions \cite{H52,B52c} about its meaning that 
continue until now \cite{Re17,Re23}. From this equation, taking into account Eq.~(\ref{9}), it follows a quantum analog of Euler's equation, which determines 
the velocity $u$,
\begin{equation}\label{12}
(\partial _{t} + u \partial _{x} ) m u = - \partial _{x} (V - Q).
\end{equation}
Equations (\ref{11}) and (\ref{12}) describe the motion of the fluid and include the additional force $\frac{1}{m} \partial _{x} Q$ caused by the quantum effects 
contained in the original Schr\"{o}dinger equation. After putting $\hbar = 0$ and interpreting $\rho$ as the matter density, these equation become equations of 
classical hydrodynamics, where the quantity $u$ is the velocity of a fluid at any given point $x$ of space at time $t$, without consideration of which particle 
occupies the position $x$. When $\hbar \neq 0$,  one can only talk about the velocity $u$ of a hypothetical fluid with the density $\rho$, where $\rho$ and $u$ 
are purely quantum characteristics and their connection with any physical substance is questionable. In this case, a hypothetical fluid is not perfect, since the 
additional force of quantum nature produces non-linear effects in the behavior of the velocity $u$.

For a stationary state of energy $W$, the probability current density $J$  is constant, $\partial_{x} J = 0$, and $\rho u = const$. The amplitude is
\begin{equation}\label{13}
A = \frac{const}{\sqrt{u}}.
\end{equation}
Then,
\begin{equation}\label{14}
\partial_{t} A = 0 \qquad \mbox{and} \qquad \partial_{t} S = - W,
\end{equation}
so that Eq.~(\ref{8}) reduces to a non-linear equation for the velocity $u$,
\begin{equation}\label{15}
\frac{m u^{2}}{2} + V - Q = W,
\end{equation}
where
\begin{equation}\label{16}
Q = \frac{\hbar^{2}}{2 m} \left[\frac{3}{4} \left(\frac{\partial_{x} u}{u} \right)^{2} - \frac{1}{2}\, \frac{\partial_{x}^{2} u}{u} \right].
\end{equation}
The gradient of $Q$ is
\begin{equation}\label{17}
\partial_{x} Q = \frac{\hbar^{2}}{2 m} \left[2 \frac{\partial_{x} u\, \partial_{x}^{2} u} {u^{2}} - \frac{3}{2} \left(\frac{\partial_{x} u}{u} \right)^{3} - \frac{1}{2} \frac{\partial_{x}^{3} u}{u} \right].
\end{equation}

\subsection{Semi-classical Bohm potential}\label{sec:2.2}
From dimensional reasons, it follows that
\begin{equation}\label{18}
Q(x) =  \frac{\hbar^{2}}{2 m x^{2}} \, \lambda(x),
\end{equation}
where $\lambda(x)$ is a dimensionless coupling function of $x$. As we will see later, the functions $Q(x)$ or $\lambda(x)$ can tend to a constant under 
certain conditions. We will calculate these functions in the semi-classical approximation. Since
$Q \sim \hbar^{2}$, then in this approximation the quantum addition can be neglected in Eq.~(\ref{15}) and we can set $u = \sqrt{\frac{2}{m} (W - V)}$.
This allows us to calculate the quantum potential (\ref{16}) explicitly,
\begin{equation}\label{19}
Q =  \frac{\hbar^{2}}{2 m} \left[\frac{5}{16} \left(\frac{\partial_{x} V}{W - V} \right)^{2} + \frac{1}{4}\, \frac{\partial_{x}^{2} V}{W - V} \right].
\end{equation}
Generally speaking, if the solution to Eq.~(\ref{3}) is known in an explicit form, then the quantum Bohm potential can be restored from Eq.~(\ref{9}). But then 
the answer will only be of a informative nature, since the quantum problem is assumed to be already solved and Eq.~(\ref{8}) is reduced to identity. Applying 
the fact that the quantum potential is proportional to $\hbar^{2}$ and, according to Eq.~(\ref{8}), it is a summand added to other summands in which the Planck 
constant is absent, allows us to find the Bohm potential using perturbation theory and the explicit form of the potential $V(x)$ only.

Consider some special cases of the potential $V(x)$. 

\subsubsection{Free fall of a particle}\label{sec:2.2.1}
For the problem of free fall of a particle of mass $m$ over the earth's surface, we have the gravitation potential in its standard form $V(x) = m g x$, where
$x$ is the height over the earth's surface. Then the quantum potential (\ref{19}) takes the form
\begin{equation}\label{23}
Q = \frac{5}{32} \frac{\hbar^{2} m g^{2}}{(W - m g x)^{2}}.
\end{equation}
This equation can be reduced to Eq.~(\ref{18}) with
\begin{equation}\label{24}
\lambda (x) = \frac{5}{16} \left(\frac{m g x}{W - m g x} \right)^{2}.
\end{equation}
If $W \gg m g x$, then we have $\lambda(x) = \frac{5}{16} (\frac{x}{x_{0}})^{2}$, where $x_{0} \equiv \frac{W}{m g}$,  and the quantum potential $Q$ does 
not depend on $x < x_{0}$.

\subsubsection{Particle in a static fluid}\label{sec:2.2.2}
When a particle is immersed in a static fluid, it is subject to the buoyant force which acts in the direction opposite to the gravitational force,
so that the total potential is $V(x) = - (m - \rho_{f} V_{d}) g x$, where $x$ is is the distance to the fluid surface, $\rho_{f}$ is the density of the fluid, $V_{d}$ is 
the volume of fluid displaced. Then the coupling function in Eq.~(\ref{18}) becomes
\begin{equation}\label{25}
\lambda (x) = \frac{5}{16} \left(\frac{(m - \rho_{f} V_{d}) g x}{W + (m - \rho_{f} V_{d}) g x} \right)^{2}.
\end{equation}
For a particle moving with energy $W \ll (m - \rho_{f} V_{d}) g x$, the constant $\lambda = \frac{5}{16}$ determines the quantum potential (\ref{18}), 
so that it does not depend on the gravitational field $g$ and fluid properties.
In the opposite limit, $W \gg (m - \rho_{f} V_{d}) g x$, the situation is similar to the one discussed in \ref{sec:2.2.1}, so that the quantum potential $Q$ is 
constant for small values of $x$.

\subsubsection{Charged particle in an accelerating electrical field}\label{sec:2.2.3}
When considering the one-dimensional motion of a particle of mass $m$ and charge $- e$ in an accelerating electrical field $\mathcal{E}$, we have
$V(x) = - e \mathcal{E} x$ and the quantum potential is
\begin{equation}\label{26}
Q = \frac{5}{32} \frac{\hbar^{2}}{m} \left(\frac{e \mathcal{E}}{W + e \mathcal{E} x} \right)^{2}.
\end{equation}
In the low-energy limit, $W \ll e \mathcal{E} x$, the quantum potential (\ref{18}) is determined by the same value $\lambda = \frac{5}{16}$ as in \ref{sec:2.2.2}. 
Here, the quantum potential $Q$ does not depend on the electrical field $\mathcal{E}$.

\subsubsection{Harmonic oscillator}\label{sec:2.2.4}
For the harmonic oscillator with the angular frequency $\omega$, the  quantum potential (\ref{19}) reduces to
\begin{equation}\label{27}
Q = \frac{\hbar^{2}}{2} \frac{\omega^{2}}{2 W - m \omega^{2} x^{2}} \left[\frac{5}{4}\, \frac{m \omega^{2} x^{2}}{2 W - m \omega^{2} x^{2}} + \frac{1}{2}\right].
\end{equation}
At high energies, $W \gg \frac{m}{2} \omega^{2} x^{2}$, the coupling function approaches the value $\lambda(x) = \frac{1}{2} (\frac{x}{x_{0}})^{2}$,
where $x_{0} \equiv \sqrt{\frac{2 W}{m \omega^{2}}}$, and $Q(x) = \frac{\hbar^{2} \omega^{2}}{8 W}$. 

\subsubsection{Square potential well}\label{sec:2.2.5}
Additionally, we will consider the square potential well: $V(x) = - V_{0}$, if $x < R$, and $V(x) = 0$, if $x > R$, where $x \geq 0$. Substituting this potential into
Eq.~(\ref{19}) gives the quantum potential
\begin{equation}\label{28}
Q = - \frac{\hbar^{2}}{8 m} \frac{V_{0}}{W}\, \frac{\delta (x - R)}{x - R}.
\end{equation}
This potential is singular. Therefore, the direct physical meaning has a mean value $\langle Q \rangle$ with respect to the states $\psi$,
\begin{equation}\label{29}
\langle Q \rangle = - \frac{\hbar^{2}}{8 m} \frac{V_{0}}{W}\, \partial_{R} \rho(R),
\end{equation}
where the density $\rho(R) = |\psi(R)|^{2}$. 

In the case of bound states with energies $W = - \frac{\hbar^{2} \kappa^{2}}{2 m} < 0$, the energy density is $\rho(R) =(R + \frac{1}{\kappa})^{-1}$ and
\begin{equation}\label{30}
\langle Q \rangle = \frac{V_{0}}{4} \frac{1}{(\kappa R +1)^{2}}.
\end{equation}

\subsubsection{Generalized case}\label{sec:2.2.6}
Let $V(x) = \Omega_{n} x^{n}$ (there is no summation over $n$ in this expression), where 
the coupling constant $\Omega_{n}$ has a dimension of Energy / Length$^{n}$. Then the quantum potential takes the form (\ref{18}) with
\begin{equation}\label{20}
\lambda(x) = \frac{\Omega_{n} x^{n}}{W - \Omega_{n} x^{n}} \left[\frac{5}{16} n^{2} \frac{\Omega_{n} x^{n}}{W - \Omega_{n} x^{n}} + 
\frac{1}{4}   n (n - 1)\right].
\end{equation}

It should be pointed out that at low energies the parameters $\Omega_{n}$ of the potential $V(x)$ do not affect the quantum 
potential $Q$. Indeed, for the motion of a particle with the energy $W \ll \Omega_{n} x^{n}$, the quantum potential takes the form (\ref{18}) with
\begin{equation}\label{21}
\lambda = \frac{1}{4} n \left(\frac{1}{4} n + 1 \right).
\end{equation}
In the reverse limit $W \gg \Omega_{n} x^{n}$, the coupling function $\lambda(x)$ tends to
\begin{equation}\label{22}
\lambda (x) = \frac{5}{16} n^{2} \left(\frac{x}{x_{0}} \right)^{2 n} + \frac{1}{4} n (n - 1) \left(\frac{x}{x_{0}} \right)^{n},
\end{equation}
where $x_{0} \equiv (W / \Omega_{n})^{1/n}$.

The particular cases $n = 1$ and $n = 2$ are considered above.

\section{Quantum geometrodynamics in $(1+1)$ dimensions}\label{sec:3}
\subsection{Basic equations}\label{sec:3.1}
A minisuperspace model with a finite number of degrees of freedom may provide a reasonable framework for addressing the problems of quantum gravity. 
In this paper, we consider the homogeneous and isotropic quantum cosmological system (further referred to as the universe), whose geometry is determined 
by the Robertson--Walker line element with the cosmic scale factor $a$. It is assumed that such a universe is originally filled with a uniform matter field $\phi$ 
and a reference perfect fluid. Following Dirac's scheme of canonical quantization, the basic equations describing such a quantum universe in Planck units can 
be reduced to the form \cite{K02,K08,K13,K18,K24}
\begin{equation}\label{31}
\left(- i \partial_{T} - \frac{2}{3} E \right) \Psi = 0,
\end{equation}
\begin{equation}\label{32}
\left(- i \partial_{a}^{2} + \kappa a^{2} - 2 a H_{\phi} - E \right) \Psi = 0,
\end{equation}
where $\Psi = \Psi (a, \phi; T)$ is a state vector, $T$ is the conformal time connected with the proper time $t$ by the differential equation $dt = a dT$, and 
$dT = N d \eta$, $N$ is the lapse function whose choice is arbitrary \cite{ADM62}, $\eta$ is the ``arc time''. The parameter $\kappa = +1, 0, -1$ is the curvature 
parameter, and $H_{\phi}$ is the Hamiltonian of the matter field $\phi$. For example, for the uniform scalar field 
\begin{equation}\label{33}
H_{\phi} = \frac{a^{3}}{2} \rho_{\phi}, \quad \rho_{\phi} = - \frac{2}{a^{6}} \partial_{\phi}^{2} + V(\phi),
\end{equation}
where $\frac{a^{3}}{2} = \mathcal{V}$ is a proper volume, $\rho_{\phi}$ is the energy density, and $V(\phi)$ is the potential. Further on, all quantities will remain 
dimensionless.

The parameter $E$ is determined by the energy density of relativistic matter (radiation) $\rho_{\gamma}$ according to the equation 
$E = a^{4} \rho_{\gamma}$, and defines a matter reference frame used to track time $T$ \cite{K02,K08,Ku91,BM96}. 
Note that in standard physical units $E$ and $T$ have the dimensions: $[E] = \mbox{Energy} \times \mbox{Length}$, $[T] = \mbox{Radians}$.

Using Eq.~(\ref{31}), Eq.~(\ref{32}) can be rewritten as a time-dependent equation in $(2+1)$ dimensions
\begin{equation}\label{34}
- i \frac{3}{2} \partial_{T} \Psi = \left(- \partial_{a}^{2} + \kappa a^{2} - 2 a H_{\phi} \right) \Psi.
\end{equation}
Let us introduce a complete orthonormal set of functions $| \Phi_{i} \rangle$, $\sum_{i} | \Phi_{i} \rangle \langle \Phi_{i} | = 1$, 
$\langle \Phi_{i} | \Phi_{k} \rangle = \delta_{i k}$, that diagonalizes the Hamiltonian $H_{\phi}$, 
\begin{equation}\label{35}
\langle \Phi_{i} | H_{\phi} | \Phi_{k} \rangle = M_{i}(a) \delta_{i k},
\end{equation}
where $M_{i}(a)$ is the mass-energy of matter in the universe in the discrete and/or continuous $i$th state, obtained after the averaging of $H_{\phi}$ with  
respect to the field $| \Phi_{i} \rangle$ in a proper volume $\frac{a^{3}}{2}$. In general case, the mass-energy of matter may depend on $a$. When $H_{\phi}$ 
describes the homogeneous scalar field, the matter turns into a barotropic fluid \cite{K18}.

Expanding the vector $\Psi$ in a series with respect to the functions $| \Phi_{i} \rangle$, 
\begin{equation}\label{36}
\Psi = \sum_{i} | \Phi_{i} \rangle \langle \Phi_{i} | \Psi \rangle,
\end{equation}
and substituting it into Eq.~(\ref{34}), we pass to the equation in $(1+1)$ dimensions
\begin{equation}\label{37}
- i \frac{3}{2} \partial_{T} \psi = \left[- \partial_{a}^{2} + U(a) \right] \psi,
\end{equation}
where $\psi = \psi (a, T) \equiv \langle \Phi_{i} | \Psi \rangle$, and
\begin{equation}\label{37a}
U(a) = \kappa a^{2} - 2 a M(a)
\end{equation}
plays the role of a potential. The state of matter index $i$
will remain unchanged and may be omitted hereafter. Equation (\ref{37})  is similar to the Schr\"{o}dinger equation (\ref{3}) except that the time variable enters 
it with the opposite sign due to the specifics of general relativity.

The solution to the equation will be found in polar form
\begin{equation}\label{38}
\psi (a, T) = A(a, T)\,e^{i S(a, T)},
\end{equation}
where the amplitude $A$ and the phase $S$ are real functions. Then Eq.~(\ref{37}) reduces to two equations
\begin{equation}\label{39}
- \frac{3}{4} \partial_{T} A^{2} + \partial_{a} \left(A^{2} \partial_{a} S \right) = 0,
\end{equation}
\begin{equation}\label{40}
- \frac{3}{2} \partial_{T} S + \left(\partial_{a} S \right)^{2} + U = Q,
\end{equation}
where
\begin{equation}\label{41}
Q = \frac{\partial_{a}^{2}  A}{A}
\end{equation}
is an analog of the quantum Bohm potential (\ref{9}). It can be considered as an additional source generated by the amplitude of the wave function.

One may consider a test particle that moves along with an idealized homogeneous fluid that represents the matter in the universe smeared out. As a result of 
the expansion of the universe as a whole, a particle acquires the velocity consistent with Hubble's law. It is preferable to consider the cosmological system 
(universe) on its own as a ``particle'' or an infinitesimal volume element that moves with the velocity $v = \frac{d a}{d T}$ in the minisuperspace. It can be 
shown \cite{K24} that the velocity $v$ is connected with the derivative of the phase $\partial_{a} S$ by a simple relation
\begin{equation}\label{42}
v = - \partial_{a} S.
\end{equation}
Then Eqs.~(\ref{39}) and (\ref{40}) take the form of hydrodynamic equations
\begin{equation}\label{43}
\frac{3}{4}\partial_{T} \rho + \partial_{a} (\rho v) = 0,
\end{equation}
\begin{equation}\label{44}
\left(\frac{3}{4} \partial_{T} + v \partial_{a} \right) v = - \frac{1}{2}\partial _{a} (U - Q),
\end{equation}
where $\rho = A^{2} = | \psi |^{2}$ is the probability density, and $\rho v$ is the density of the probability current of an infinitesimal volume element at point $a$ 
and at time $T$. Equations (\ref{43}) and (\ref{44}) are a continuity equation and a quantum analog of Euler's equation, respectively. 

The generalized force 
\begin{equation}\label{44a}
\mathcal{F} = - \frac{1}{2}\partial _{a} (U - Q)
\end{equation}
can be calculated explicitly,
\begin{equation}\label{44b}
\mathcal{F} = - a \kappa + \frac{a^{3}}{2} \left(\rho_{m} - 3 p_{m} \right) + \frac{1}{2} \partial_{a} Q,
\end{equation}
where $\rho_{m} = \frac{M}{\mathcal{V}}$ is the matter energy density, $p_{m} = - \frac{d M}{d \mathcal{V}}$ is the pressure.
The perfect nature of the fluid is broken by the quantum term. If this term is neglected, Eqs.~(\ref{43}) and (\ref{44}) acquire only a formal resemblance to the equations of classical hydrodynamics, since $\rho$ is not the density of real matter, but the probability density.

\subsection{Stationary state}\label{sec:3.2}
Substituting Eq.~(\ref{38}) into Eq.~(\ref{31}), we obtain the condition on the wave function $\psi \neq 0$,
\begin{equation}\label{45}
\left(- i \frac{\partial_{T} A}{A} + \partial_{T} S - \frac{2}{3} E \right) \psi = 0.
\end{equation}
This leads to two equations,
\begin{equation}\label{46}
\partial_{T} A = 0 \quad \mbox{and} \quad \partial_{T} S = \frac{2}{3} E.
\end{equation}
They are similar to Eq.~(\ref{14}) and describe a ``stationary state'' with a given parameter $E$. The requirement of 
``stationarity'' is not additionally imposed on the quantum system, but it follows from the equations of
quantum geometrodynamics (\ref{31}) and (\ref{32}). Then, from Eqs.~(\ref{39}) and (\ref{40}), we come to the equations
\begin{equation}\label{47}
\partial_{a} (\rho v) = 0,
\end{equation}
\begin{equation}\label{48}
v^{2} + U - E - Q = 0.
\end{equation}
From Eq.~(\ref{47}), we find the probability density
\begin{equation}\label{49}
\rho = \frac{const}{v},
\end{equation}
so that the quantum potential takes the form
\begin{equation}\label{50}
Q =  \frac{3}{4} \left(\frac{\partial_{a} v}{v} \right)^{2} - \frac{1}{2}\, \frac{\partial_{a}^{2} v}{v},
\end{equation}
and its derivative with respect to $a$ becomes
\begin{equation}\label{44c}
\partial_{a} Q = 2 \frac{\partial_{a} v\, \partial_{a}^{2} v} {v^{2}} - \frac{3}{2} \left(\frac{\partial_{a} v}{v} \right)^{3} - \frac{1}{2} \frac{\partial_{a}^{3} v}{v}.
\end{equation}
A positive quantum potential leads to faster expansion of the universe.

\subsection{Semi-classical examples}\label{sec:3.3}
Since in standard physical units, the potential $Q \sim \hbar^{2}$ \cite{K13} (cf. Eq.~(\ref{16})), then Eq.~(\ref{48}) can be solved via perturbation theory.
In the Born approximation, we obtain
\begin{equation}\label{51}
Q =  \frac{5}{16} \left(\frac{\partial_{a} U}{E - U} \right)^{2} + \frac{1}{4}\, \frac{\partial_{a}^{2} U}{E - U}.
\end{equation}
Let's take a look at a few specific examples.

\subsubsection{Radiation dominance}\label{sec:3.3.1}
For such a universe, $E \neq 0$, $M = 0$, and $\kappa = 0$. Then $U = 0$, and the quantum addition $Q = 0$. This result is consistent with the one obtained before in Ref.~\cite{K24} under other assumptions.

\subsubsection{Matter dominance}\label{sec:3.3.2}
In the case, $E = 0$, $M \neq 0$, and $\kappa = 0$, we have $U = - 2 a M$, and
\begin{equation}\label{52}
Q = \frac{5}{16} \frac{1}{a^{2}},
\end{equation}
if matter is dust, and 
\begin{equation}\label{53}
Q =  \frac{\lambda(a)}{a^{2}},
\end{equation}
where the coupling function is
\begin{equation}\label{54}
\lambda(a) = \frac{5}{16} \left(1 + a \frac{\partial_{a} M}{M} \right)^{2} + \frac{a}{4} \frac{\partial_{a} M}{M} \left[2 + a \frac{\partial_{a}^{2} M}{\partial_{a} M} \right],
\end{equation}
if $\partial_{a} M \neq 0$, and $\partial_{a}^{2} M \neq 0$. The derivatives of $M$ define the pressure 
\begin{equation}\label{55}
p_{m} = - \frac{2}{3 a^{2}} \partial_{a} M
\end{equation}
and its derivative
\begin{equation}\label{56}
\partial_{a} p_{m} = \frac{4}{3 a^{3}} \partial_{a} M - \frac{2}{3 a^{2}} \partial_{a}^{2} M.
\end{equation}

\subsubsection{Spatially flat universe with radiation and dust}\label{sec:3.3.3}
For the universe with $\kappa  = 0$, $E \neq 0$, and $M = const$, we get $U = - 2 a M$ again, and the quantum potential is
\begin{equation}\label{57}
Q = \frac{5}{4} \left(\frac{M}{E + 2 a M} \right)^{2}.
\end{equation}
It reduces to Eq.~(\ref{52}) in the case $E = 0$, and we have $Q = \frac{5}{4} \left( \frac{M}{E} \right)^{2}$ at the point $a = 0$, so that the quantum potential 
is non-singular.

\subsubsection{Empty universe with spatial curvature}\label{sec:3.3.4}
If $\kappa \neq 0$, $E = 0$, and  $M = 0$, we obtain $U = \kappa a^{2}$ and
\begin{equation}\label{58}
Q = \frac{3}{4} \frac{1}{a^{2}}.
\end{equation}
We see that spatial curvature itself can generate quantum potential, even in an empty universe, with no dependence on the curvature parameter.

\subsection{Quantum potential impact on expansion}\label{sec:3.4}
Equation (\ref{48}) can be rewritten in the form convenient for the comparison with general relativity
\begin{equation}\label{59}
\left(\frac{v}{a^{2}}\right)^{2} = \rho_{\mbox{\tiny mat}} + \rho_{\mbox{\tiny B}},
\end{equation}
where 
\begin{equation}\label{60}
\rho_{\mbox{\tiny mat}} = \frac{2 M}{a^{3}} + \frac{E}{a^{4}} - \frac{\kappa}{a^{2}}, \quad \rho_{\mbox{\tiny B}} = \frac{Q}{a^{4}},
\end{equation}
$\rho_{\mbox{\tiny mat}}$ is the total energy density with the contributions from matter with mass $M$, radiation, and the term accounting for the curvature of space;
$\rho_{\mbox{\tiny B}}$ is the energy density stipulated by the quantum Bohm potential. As we can see, taking into account the quantum potential can affect 
the Hubble expansion rate $H = \frac{v}{a^{2}}$ \cite{K24}. Equation (\ref{59}) can be viewed as the quantum analog of the first Friedmann equation, 
whereas Eq.~(\ref{44}) is the equation for acceleration $\frac{d^{2} a}{d T^{2}}$ with a quantum correction and it coincides with the corresponding equation of a 
homogeneous, isotropic model \cite{K13}.

For illustrative purposes, we will do a few calculations. 

In the simplest case of an empty universe, the quantum addition has a form (\ref{58}). By integrating Eq.~(\ref{48}), we obtain the following 
relations,
\begin{equation}\label{61}
e^{2 T} = \left|a^{2} + \sqrt{a^{4} + \frac{3}{4}}\right|, \quad \mbox{for} \quad \kappa = 1,
\end{equation}
\begin{equation}\label{62}
\begin{split}
T & = \frac{1}{2} \arcsin \frac{2}{\sqrt{3}} a^{2}, \quad \mbox{if} \quad a < \left(\frac{3}{4} \right)^{\frac{1}{4}}, \\
e^{2 i T} & = \left|a^{2} + \sqrt{a^{4} - \frac{3}{4}}\right|, \quad \mbox{if} \quad a > \left(\frac{3}{4} \right)^{\frac{1}{4}} \quad \mbox{for} \quad \kappa = - 1,
\end{split}
\end{equation}
For values $a \gg 1$, Eqs.~(\ref{61}) and (\ref{62}) reduce to
\begin{equation}\label{63}
a = \frac{e^{T}}{\sqrt{2}} \quad \mbox{for} \quad \kappa = 1, \quad \mbox{and} \quad a = \frac{e^{i T}}{\sqrt{2}} \quad \mbox{for} \quad \kappa = - 1.
\end{equation}
The same expressions follow from the equations of general relativity, if we do not take quantum effects into account.

For a spatially flat universe filled with dust and radiation, the quantum additive has a form (\ref{57}). Then from Eq.~(\ref{48}), it follows
\begin{equation}\label{64}
\alpha\, T = \frac{2 \sqrt{x^{3} + 1}}{x + 1 + \sqrt{3}} - \frac{\sqrt{3} - 1}{3^{\frac{1}{4}}}\, F(\varphi, k) - 2 \sqrt{3}\, E (\varphi, k),
\end{equation}
where
\begin{equation}\label{65}
\alpha = \left(\frac{4 M}{5^{\frac{1}{4}}} \right)^{\frac{2}{3}}, \quad x = \frac{2 a M + E}{\left(\frac{5}{4} M^{2}\right)^{\frac{1}{3}}},
\end{equation}
while $F(\varphi, k)$ and $E(\varphi, k)$ are elliptic integrals of the first and second kind with
\begin{equation}\label{66}
\varphi = \frac{x + 1 - \sqrt{3}}{x + 1 + \sqrt{3}}, \quad k = \frac{\sqrt{2 + \sqrt{3}}}{2}.
\end{equation}

\section{Discussion}\label{sec:4}
In this paper, we draw an analogy between non-relativistic quantum mechanics in the Madelung formulation and quantum 
geometrodynamics for a homogeneous and isotropic cosmological system. In quantum mechanics, one may pass from studying a particle moving in a force 
field  to describing an ensemble of identical particles. The Schr\"{o}dinger equation can be reduced to two hydrodynamic-type equations (\ref{11}) and (\ref{12}), one 
of which is the standard continuity equation and the other is the Euler equation for the velocity of the fluid subjected to an additional force due to the quantum 
Bohm potential (\ref{9}). We calculate this additional force for a number of example potentials that determine the motion of the fluid and obtain the coordinate 
dependence of the quantum potential.

The availability of a region for the fluid motion, where the quantum potential is non-vanishing, means that, according to Eq.~(\ref{12}), the motion of the fluid 
can differ considerably from that suggested by classical hydrodynamics. In particular, a quantum addition can be responsible for the diffraction pattern when a 
particle is scattered on two slits \cite{Hi79}. Basically, the quantum potential can be given by Eq.~(\ref{18}), where the singularity at small coordinate values is 
singled out, and the dimensionless coupling function, strictly speaking, retains its dependence on the coordinate. From the considered examples, it can be 
concluded, in particular, that if the classical potential is negatively defined, then for large values of the coordinate, the coupling function becomes constant, so 
that the quantum potential is inversely proportional to the square of the coordinate. For small values of the coordinate, the quantum potential itself can go to a 
constant. Only in the region, where the gradient from the quantum potential vanishes, it does not affect the velocity of the fluid.

In quantum geometrodynamics, we begin with equations (\ref{31}) and (\ref{32}) that describe a homogeneous and isotropic universe. A state vector 
$\Psi (a, \phi; T)$ is considered as a function of the cosmic scale factor $a$ that determines the geometry, a uniform scalar field $\phi$ responsible for matter in 
the universe, and a parameter $T$ which describes the evolution of a system in conformal time. These two equations can be reduced to a single 
time-dependent equation (\ref{34}) in $(2+1)$ dimensions. After averaging over the states of the material field, Eq.~(\ref{34}) is transformed into Eq.~(\ref{37}) 
in $(1+1)$ dimensions. Then, using a direct analogy with quantum mechanics in Madelung's formalism, the latter equation can be rewritten as two equations 
(\ref{39}) and (\ref{40}) for the amplitude of the wave function and its phase. The quantum potential (\ref{41}) that emerges here is completely analogous to the 
quantum Bohm potential (\ref{9}), except that the coordinate is substituted by the scale factor. In contrast to quantum mechanics, where the force field $V(x)$ 
is determined by the problem statement, in geometrodynamics the form of the potential function (\ref{37a}) is specific. After all the steps, the equations 
(\ref{39}) and (\ref{40}) can finally be reduced to two hydrodynamic-type equations (\ref{43}) and (\ref{44}).

In the hydrodynamic approach, the flow velocity $v = \frac{d a}{d T}$ is the fundamental quantity through which the Hubble expansion rate 
$H = \frac{1}{a^{2}} \frac{d a}{d T}$  is expressed.

The generalized force $\mathcal{F}$ (\ref{44b}), as in the case of quantum mechanics (see Eq.~(\ref{12})), contains an additional force stipulated by the 
quantum potential gradient, which makes the corresponding equations (\ref{44}) nonlinear with respect to the velocity field $v$. Similar to quantum mechanics, 
where nonlinear effects cause a diffraction pattern, the corresponding interference of classical trajectories should also occur in quantum geometrodynamics, 
but the consequences for observations remain unclear for now.

The transition to the so-called stationary states deserves a special mention. In contrast to quantum mechanics, the requirement of the state of the Universe 
being stationary is not imposed on the quantum system as an additional condition, but follows from the equations of quantum geometrodynamics themselves.

The explicit forms of the quantum potential in the leading Born approximation are given in Eqs.~(\ref{52}) -- (\ref{58}) for the universes dominated by one or 
another type of matter.

The impact of the quantum potential $Q$ on the expansion of the universe is demonstrated in the form of explicit solutions (\ref{61}) -- (\ref{63}) of 
Eq.~(\ref{48}) in the case of an empty universe with different spatial curvatures. The solution for a spatially flat universe filled with dust and radiation
is given by Eq.~(\ref{64}).

\begin{figure}[t]
\centering
\includegraphics[width=12cm]{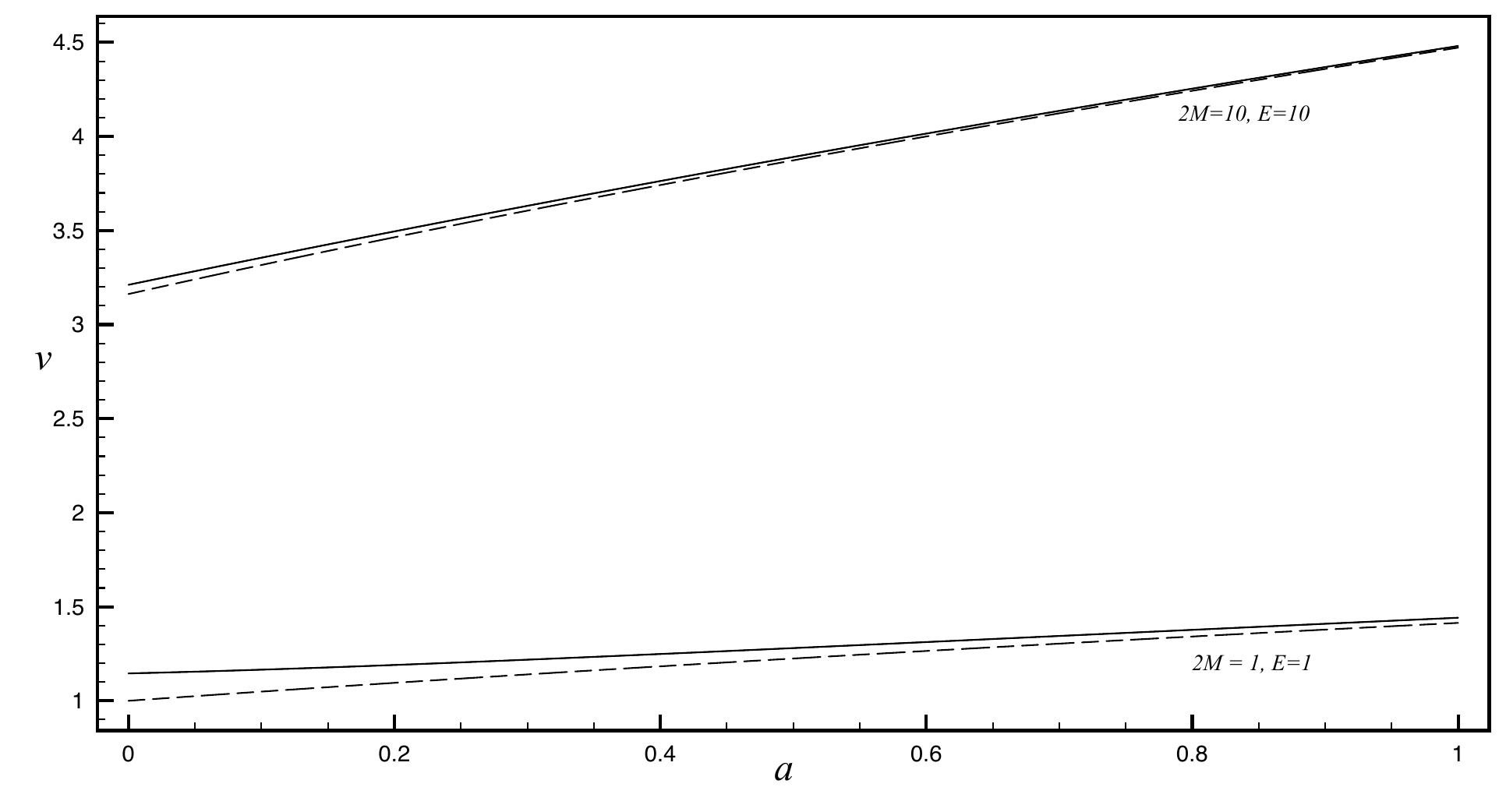}
\caption{The velocity $v$ as a function on the scale factor $a$ for different values of the parameters $M$ and $E$. Dotted lines correspond to the cases 
$Q = 0$.}
\label{fig1}
\end{figure} 

\begin{figure}[t]
\centering
\includegraphics[width=12cm]{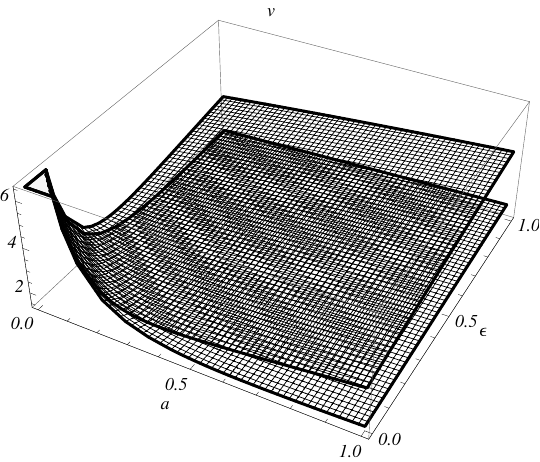}
\caption{The velocity $v$ as a function on the scale factor $a$ and the parameter $\varepsilon = \frac{E}{2 M}$.
The lower surface corresponds to the value $2 M = 2$, and the upper surface to $2 M = 10$.}
\label{fig2}
\end{figure}

\begin{figure}[t]
\centering
\includegraphics[width=12cm]{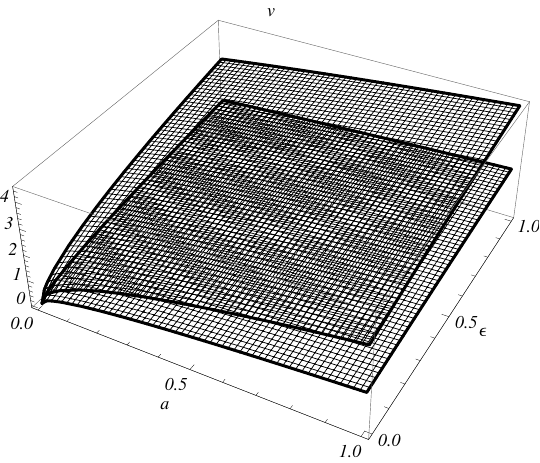}
\caption{The velocity $v$ as a function on the scale factor $a$ and the parameter $\varepsilon = \frac{E}{2 M}$ for different values of $M$ and $Q = 0$. See the caption to Fig.~\ref{fig2}.}
\label{fig3}
\end{figure}

The flow velocity $v$ as a function on $a$ is shown in Fig.~\ref{fig1} for the spatially flat universe filled with dust and radiation. The 
contribution of the quantum potential (\ref{57}) to the evolution of the universe is noticeable only at small values of $a$. It is worth noting that in this model the flow velocity is not singular at the point $a = 0$.
A three-dimensional picture for the function $v = v(a, \varepsilon)$, where $\varepsilon = \frac{E}{2 M}$, is shown in Fig.~\ref{fig2}\footnote{Different values 
$\varepsilon$ for a constant contribution of non-relativistic matter $M$ correspond to universes with different radiation densities.}. This figure is for illustrative 
purposes only. Each point on the surfaces corresponds to one particular universe with specific parameter values. The set of universes has 
a singularity $v \rightarrow \infty$, when $a = 0$ and $\varepsilon \rightarrow 0$. For comparison, Fig.~\ref{fig3} represents the case when there are no quantum effects. 

When the universe is filled with dust and radiation, from Eqs.~(\ref{59}) and (\ref{57}), we obtain
\begin{equation}\label{67}
v = a^{2} \sqrt{\rho_{m} \left(1 + \frac{\rho_{\gamma}}{\rho_{m}} \right) + \frac{5}{16} \frac{1}{a^{6}} \left(1 + \frac{\rho_{\gamma}}{\rho_{m}} \right)^{-2}}.
\end{equation}
At the epoch, when matter dominates over radiation and $\frac{\rho_{\gamma}}{\rho_{m}}  \ll 1$, the flow velocity $v$ is
\begin{equation}\label{68}
v = a^{2} \sqrt{\rho_{m}  + \frac{5}{16} \frac{1}{a^{6}}}.
\end{equation}
At the epoch, when radiation prevails over matter and $\frac{\rho_{\gamma}}{\rho_{m}}  \gg 1$, we have
\begin{equation}\label{69}
v = a^{2} \sqrt{\rho_{\gamma}  + \frac{5}{16} \frac{1}{a^{6}}\left(\frac{\rho_{m}}{\rho_{\gamma}} \right)^{2}}.
\end{equation}

In the region of values of $a$, where quantum effects can be neglected in Eq.~(\ref{68}), we pass to the classical limit, where general relativity is applicable.
When matter dominates over radiation, we get
\begin{equation}\label{70}
\frac{v}{a^{2}} = \sqrt{\rho_{m}},
\end{equation}
where $\frac{v}{a^{2}} = H$ is a Hubble expansion rate. 

If we introduce a cosmological constant $\Lambda$ with energy density $\rho_{\Lambda}$ into the model, Eq.~(\ref{70}) takes the form
\begin{equation}\label{71}
\frac{v}{a^{2}} = \sqrt{\rho_{m} + \rho_{\Lambda}}.
\end{equation}

For the universe with $\rho_{\Lambda} \approx \rho_{m}$, comparing Eqs.~(\ref{70}) and (\ref{71}), we see that the density associated with the 
cosmological constant increases the expansion rate by a factor of $\sqrt{2}$.

The case of the dominance of radiation over matter requires a more careful analysis. It is possible that the quantum term in Eq.~(\ref{69}) cannot be neglected, 
despite the fact that it is proportional to the second power of a small value $\frac{\rho_{m}}{\rho_{\gamma}}$, and everything depends on how close we approach the 
initial cosmological singularity $a = 0$. From Eq.~(\ref{69}), it follows that both terms under the square root can be estimated as being proportional to $a^{-4}$,
and $v$ tends to a constant.

\section*{Acknowledgements}
This work was partially supported by The National Academy of
Sciences of Ukraine (Projects No.~0121U109612 and  No.~0122U000886) and by a grant from the Simons Foundation (USA).

\end{document}